# Stacking-engineered ferroelectricity in bilayer boron nitride

Kenji Yasuda[1]*, Xirui Wang[1], Kenji Watanabe[2], Takashi Taniguchi[2], Pablo Jarillo-Herrero[1]*

[1] Department of Physics, Massachusetts Institute of Technology, Cambridge, Massachusetts 02139, USA.

[2] National Institute for Materials Science, Namiki 1-1, Tsukuba, Ibaraki 305-0044, Japan.

*Correspondence to: yasuda@mit.edu, pjarillo@mit.edu

**2D ferroelectrics with robust polarization down to atomic thicknesses provide novel building blocks for functional heterostructures. Experimental reports, however, remain scarce because of the requirement of a layered polar crystal. Here, we demonstrate a rational design approach to engineering 2D ferroelectrics from a non-ferroelectric parent compound via employing van der Waals assembly. Parallel-stacked bilayer boron nitride is shown to exhibit out-of-plane electric polarization that reverses depending on the stacking order. The polarization switching is probed via the resistance of an adjacently-stacked graphene sheet. Furthermore, twisting the boron nitride sheets by a small-angle changes the dynamics of switching due to the formation of moiré ferroelectricity with staggered polarization. The ferroelectricity persists to room temperature while keeping the high mobility of graphene, paving the way for potential ultrathin nonvolatile memory applications.**

Ferroelectric materials with an electric-field switchable polarization offer a wide range of technological applications, such as nonvolatile memories, high-permittivity dielectrics, electro-mechanical actuators, and pyroelectric sensors (*1*). Thinning down vertical ferroelectrics is one of the essential steps for the implementation of ferroelectric nonvolatile memory due to the quest for denser storage and lower power consumption (*1*). Room-temperature ferroelectricity down to atomic thicknesses was, however, difficult to access due to the depolarization effect until the recent development of three series of materials: epitaxial perovskites (*2*, *3*), $HfO_2$-based ferroelectrics (*4*), and low-dimensional van der Waals (vdW) ferroelectrics (*5–13*). Among them, 2D vdW ferroelectrics present unique opportunities to integrate high mobility materials such as graphene into ferroelectric field-effect transistors (FeFETs) while keeping their properties intact due to the absence of dangling bonds (*14*). Their uniform atomic thickness also makes them ideal as ferroelectric tunnel barriers for their use in ferroelectric tunnel junctions (FTJ) (*15*). Despite the potential importance for the application as a ferroelectric memory, only a few examples of 2D vertical ferroelectrics, $CuInP_2S_6$, $In_2Se_3$, $MoTe_2$, and $WTe_2$, have been discovered so far (*9–13*). Here, the candidate materials are largely limited by the requirement of the polar space group in the original layered bulk crystal.

The development of vdW assembly enabled the engineering of heterostructures with physical properties beyond the sum of those of the individual layers (*16*). For example, the Dirac band structure of graphene is dramatically transformed when it is aligned with hexagonal boron nitride (BN) or stacked with another slightly rotated graphene sheet. The modified band structures have led to the discovery of a variety of emergent phenomena related to electron correlations and topology beyond expectations from the original band structure (*17–24*). In this paper, we demonstrate that the vdW stacking modifies not only the electronic band structure but also the crystal symmetry, thereby enabling the design of ferroelectric materials out of non-ferroelectric parent compounds. Here, we use boron nitride (BN) as an example, but the same procedure can be applied to other bipartite honeycomb 2D materials, such as 2H-type transition metal dichalcogenides (TMDs) (*25*). Bulk hexagonal BN (hBN) crystals realize AA' stacking, as displayed in Fig. 1A. This 180° rotated natural stacking order restores the inversion symmetry broken in the monolayer. However, if two hBN monolayer sheets are stacked without rotation (parallel stacking, P), noncentrosymmetric AB or BA stacking orders (Fig. 1, B and C, respectively) are energetically favored (*26*). In these configurations, the B (N) atoms in the upper layer sit above the N (B) atoms in the lower layer, while the N (B) atoms in the upper layer lay above the empty site at the center of the hexagon in the lower layer. The vertical alignment of the $2p_z$ orbitals of N and B distorts the orbital of N, creating an electric dipole moment (fig. S2). The opposite dipole moment for AB and BA stacking presents the interesting possibility that the out-of-plane polarization can be switched by an in-plane interlayer shear motion of one-third of the unit cell (*25*).

Experimentally, we form nearly 0° bilayer hBN devices by using the "tear and stack" method, where half of a monolayer BN flake is picked up, and stacked on top of the remaining half (*27*, *28*). To study the change of the polarization under an electric field, we fabricated dual gated vdW heterostructure devices composed of metal top gate (Au/Cr)/hBN/graphene/0° parallel stacked bilayer BN (P-BBN)/hBN/metal bottom gate (PdAu) (e.g. device P1), as schematically shown in the inset of Fig. 1D. Here the graphene sensitively detects the extra charge carriers induced by the polarization of P-BBN. Figure 1D shows the resistance of the graphene sensor as a function of top gate, $V_T$ (for both forward and backward gate sweep directions), which exhibits a typical maximum without hysteresis. In contrast, the forward and backward scans of the resistance versus the bottom gate voltage, $V_B$ (Fig. 1E) shows hysteresis, exhibiting maxima at about 0.10 V/nm and 0.12 V/nm for the backward and forward scans, respectively. In addition, we observe a resistance step at around 0.20 V/nm in the forward scan, as displayed in the inset. As discussed below, this bistability is attributed to the polarization switching of P-BBN by the applied electric field.

Dual-gate scanning allows independent control of the carrier density of graphene and the electric field across the P-BBN, since the top gate primarily changes the former (figs. S4 and S5) while the bottom gate changes both. In a standard dual-gated graphene device, a measurement of the resistance versus top and bottom gate voltages results in a single diagonal feature, a maximum resistance ridge, corresponding to the charge neutrality condition. The diagonal feature stems from the fact that the induced carrier density follows the equation $n=\varepsilon_{hBN}(V_B/d_B+V_T/d_T)$, where $\varepsilon_{hBN}$ is the dielectric constant of hBN, and $d_B$ ($d_T$) is the distance between graphene and the bottom (top) gate electrode. In contrast, two parallel-shifted diagonal lines are observed in a dual-gate scan for our P-BBN device (Fig. 2A). The shift reflects an abrupt change in the induced carrier density, $\Delta n_P$, caused by the switching of the electric polarization of P-BBN: As the polarization switches from up (BA stacking) to down (AB stacking) at $V_B/d_B = -0.06$ V/nm, the total induced carrier density changes from $\varepsilon_{hBN}(V_B/d_B+V_T/d_T)+\Delta n_P$ to $\varepsilon_{hBN}(V_B/d_B+V_T/d_T)-\Delta n_P$, leading to the shift of the charge neutrality resistance peak. Similarly, the forward scan of the bottom gate shows the polarization switching from down to up at $V_B/d_B = 0.16$ V/nm (Fig. 2B). Interestingly, the resistance measured using the lower voltage contacts exhibits an intermediate, two-peak behavior during the switching (Fig. 2C). This indicates the coexistence of micrometer-scale AB and BA domains and provides a hint to the dynamics of polarization switching. Namely, the domain wall is pinned in the middle of the Hall bar at around $V_B/d_B = 0.13$ V/nm, followed by the depinning at around $V_B/d_B = 0.16$ V/nm.

To further investigate the ferroelectric properties of P-BBN, we measure the carrier density $n_H$ of graphene extracted from Hall resistance measurements (Fig. 2D). Hysteretic behavior with an abrupt jump of $n_H$ is observed when sweeping $V_B$, which is attributed to the ferroelectric switching. The subtraction of the forward and backward sweeps gives the magnitude of $2\Delta n_P$, which equals $3.0 \times 10^{11}$ cm$^{-2}$ (Fig. 2E). This value is consistent with $2\Delta n_P = 2.6 \times 10^{11}$ cm$^{-2}$ estimated from the horizontal shift of the charge neutrality resistance peak in the dual-gate scan (Fig. 2, A to C). $\Delta n_P$ allows us to calculate the magnitude of the polarization of AB-stacked bilayer BN. According to a simple model calculation (see fig. S8 for details), the two-dimensional polarization follows $P_{2D} = e\Delta n_P d_B$; namely, the electric dipole moment between the bottom gate and graphene is equal to the magnitude of the polarization of bilayer BN. Figure 2F shows our measurement of $2\Delta n_P$ for four different devices studied in this work, which indeed exhibit an inverse proportional behavior with respect to $d_B$. The magnitude of the polarization estimated from these data points is $P_{2D} = 2.25\ (0.37) \times 10^{-12}$ C m$^{-1}$ (corresponding to $P_{3D} = 0.68$ μC cm$^{-2}$). This agrees well with the theoretically-calculated magnitude of the polarization of AB-stacked bilayer BN from a Berry phase calculation, $P_{2D,theory} = 2.08 \times 10^{-12}$ C m$^{-1}$ (*25*, *29*).

Having established the ferroelectric nature of AB-stacked bilayer BN, we employ the unique characteristics of vdW heterostructures to control the domain structures further. When two identical layers are stacked with a small rotation angle, a moiré pattern forms with a much larger periodicity than the in-plane lattice constant. In small-angle twisted bilayer BN, AB and BA lattice networks with topological defects (AA regions) and domain walls are formed due to the lattice relaxation, like the case of twisted bilayer graphene and TMDs (Fig. 3A) (*30*, *31*). Here, due to the opposite polarization of AB and BA stacking regions, these domains with staggered polarization are expected to expand or shrink, through domain wall motion, when a vertical electric field is applied. Figure 3B shows the dual-gate scan of the resistance of graphene for a 0.6° rotation-angle twisted bilayer BN (device T1). It exhibits two parallel diagonal peaks, each corresponding to AB or BA domains, similar to Fig. 2A. However, rather than an abrupt transition between the two lines, a gradual shift in weight from one to the other takes place along the diagonal. Thus, the magnitude of each peak gives the relative proportion of AB and BA domain sizes, or the average polarization, as a function of the applied electric field (fig. S11). The electric field dependence of the polarization (Fig. 3C) highlights the difference between the twisted and non-twisted devices. Firstly, the coercive field is much smaller for the twisted bilayer BN than the non-twisted P-BBN. Secondly, the polarization switching occurs gradually, in contrast to the sharp switching of the non-twisted device. In a non-twisted device, a domain wall moves over the device scale during the switching as discussed in Fig. 2, A to C, and is likely to be pinned by strong pinning centers. In contrast, each domain wall in a twisted device moves only by a moiré length scale and will experience weaker pinning centers, leading to the small coercive field. Besides, the different pinning strength of each domain wall leads to the gradual switching. Thus, the global rotation of the two layers modifies the dynamics of the ferroelectric switching behavior.

Finally, we study the temperature dependence of the ferroelectricity in P-BBN. Interestingly, the polarization measured from $\Delta n_P$ is almost independent of temperature (Fig. 4A, fig. S16) up to room temperature. The nearly temperature-independent ferroelectric polarization presumably reflects the unique coupling between the out-of-plane polarization and in-plane shear motion in P-BBN. The strong intralayer covalent bonding inhibits the in-plane thermal vibration of atoms, making the polarization insensitive to temperature (*32*). Correspondingly, the ferroelectric hysteresis is observable up to room temperature despite the temperature-induced broadening of the resistance peak (Fig. 4B). Such hysteretic behavior allows us to deterministically write the polarization by a voltage pulse of only a few volts and read it in a nonvolatile way, as shown in Fig. 4C. We also checked the stability of the ferroelectric polarization by keeping the sample at 0 V at room temperature for an extended period after setting the polarization to up or down (Fig. 4D). The resistance remains almost the same after at least a month (the longest period measured); namely, P-BBN retains its polarization over a technologically-relevant time scale. Hence, the present result points to the potential use of P-BBN/graphene as a ferroelectric FET with ultrahigh mobility of graphene of around $5 \times 10^4$ $cm^2V^{-1}s^{-1}$ at room temperature (figs. S13 to S15).

The designer approach for engineering vdW ferroelectrics and moiré ferroelectrics demonstrated in this study can be extended to other bipartite honeycomb 2D materials, such as semiconducting 2H-type TMDs like $MoS_2$ and $WSe_2$, metallic and superconducting ones like $NbS_2$ and $NbSe_2$, and Group III chalcogenides like GaS, GaSe, and InSe (*25*). The inversion symmetry breaking of these synthetic ferroelectrics will be coupled to the electronic band structures in a tunable manner through polarization switching. In addition to novel fundamental physics resulting from the modification of the intrinsic properties of each material, such engineered ferroelectrics and moiré systems may significantly expand the capabilities of 2D materials for electronic, spintronic, and optical applications (*15*, *33*).

**References and Notes:**

1. K. Uchino, *Ferroelectric devices* (CRC press, 2009).
2. D. D. Fong, G. B. Stephenson, S. K. Streiffer, J. A. Eastman, O. Auciello, P. H. Fuoss, C. Thompson, Ferroelectricity in ultrathin perovskite films. *Science*. **304**, 1650–1653 (2004).
3. H. Wang, Z. R. Liu, H. Y. Yoong, T. R. Paudel, J. X. Xiao, R. Guo, W. N. Lin, P. Yang, J. Wang, G. M. Chow, T. Venkatesan, E. Y. Tsymbal, H. Tian, J. S. Chen, Direct observation of room-temperature out-of-plane ferroelectricity and tunneling electroresistance at the two-dimensional limit. *Nat. Commun.* **9**, 3319 (2018).
4. U. Schröeder, C. S. Hwang, H. Funakubo, *Ferroelectricity in doped hafnium oxide: materials, properties and devices* (Woodhead Publishing, 2019).
5. C. Cui, F. Xue, W.-J. Hu, L.-J. Li, Two-dimensional materials with piezoelectric and ferroelectric functionalities. *npj 2D Mater. Appl.* **2**, 18 (2018).
6. M. Wu, P. Jena, The rise of two-dimensional van der Waals ferroelectrics. *Wiley Interdiscip. Rev. Comput. Mol. Sci.* **8**, e1365 (2018).
7. A. V. Bune, V. M. Fridkin, S. Ducharme, L. M. Blinov, S. P. Palto, A. V. Sorokin, S. G. Yudin, A. Zlatkin, Two-dimensional ferroelectric films. *Nature*. **391**, 874–877 (1998).
8. K. Chang, J. Liu, H. Lin, N. Wang, K. Zhao, A. Zhang, F. Jin, Y. Zhong, X. Hu, W. Duan, Q. Zhang, L. Fu, Q.-K. Xue, X. Chen, S.-H. Ji, Discovery of robust in-plane ferroelectricity in atomic-thick SnTe. *Science*. **353**, 274–278 (2016).

9.  F. Liu, L. You, K. L. Seyler, X. Li, P. Yu, J. Lin, X. Wang, J. Zhou, H. Wang, H. He, S. T. Pantelides, W. Zhou, P. Sharma, X. Xu, P. M. Ajayan, J. Wang, Z. Liu, Room-temperature ferroelectricity in $CuInP_2S_6$ ultrathin flakes. *Nat. Commun.* **7**, 12357 (2016).

10. Y. Zhou, D. Wu, Y. Zhu, Y. Cho, Q. He, X. Yang, K. Herrera, Z. Chu, Y. Han, M. C. Downer, H. Peng, K. Lai, Out-of-plane piezoelectricity and ferroelectricity in layered α-$In_2Se_3$ nanoflakes. *Nano Lett.* **17**, 5508–5513 (2017).

11. C. Cui, W. J. Hu, X. Yan, C. Addiego, W. Gao, Y. Wang, Z. Wang, L. Li, Y. Cheng, P. Li, X. Zhang, H. N. Alshareef, T. Wu, W. Zhu, X. Pan, L. J. Li, Intercorrelated In-Plane and Out-of-Plane Ferroelectricity in Ultrathin Two-Dimensional Layered Semiconductor $In_2Se_3$. *Nano Lett.* **18**, 1253–1258 (2018).

12. S. Yuan, X. Luo, H. L. Chan, C. Xiao, Y. Dai, M. Xie, J. Hao, Room-temperature ferroelectricity in $MoTe_2$ down to the atomic monolayer limit. *Nat. Commun.* **10**, 1775 (2019).

13. Z. Fei, W. Zhao, T. A. Palomaki, B. Sun, M. K. Miller, Z. Zhao, J. Yan, X. Xu, D. H. Cobden, Ferroelectric switching of a two-dimensional metal. *Nature*. **560**, 336–339 (2018).

14. C. R. Dean, A. F. Young, I. Meric, C. Lee, L. Wang, S. Sorgenfrei, K. Watanabe, T. Taniguchi, P. Kim, K. L. Shepard, J. Hone, Boron nitride substrates for high-quality graphene electronics. *Nat. Nanotechnol.* **5**, 722–726 (2010).

15. E. Y. Tsymbal, H. Kohlstedt, Tunneling across a ferroelectric. *Science.* **313**, 181–183 (2006).

16. A. K. Geim, I. V. Grigorieva, Van der Waals heterostructures. *Nature*. **499**, 419–425 (2013).

17. B. Hunt, J. D. Sanchez-Yamagishi, A. F. Young, M. Yankowitz, B. J. LeRoy, K. Watanabe, T. Taniguchi, P. Moon, M. Koshino, P. Jarillo-Herrero, R. C. Ashoori, Massive Dirac Fermions and Hofstadter Butterfly in a van der Waals Heterostructure. *Science.* **340**, 1427–1431 (2013).

18. C. R. Dean, L. Wang, P. Maher, C. Forsythe, F. Ghahari, Y. Gao, J. Katoch, M. Ishigami, P. Moon, M. Koshino, T. Taniguchi, K. Watanabe, K. L. Shepard, J. Hone, P. Kim, Hofstadter's butterfly and the fractal quantum Hall effect in moiré superlattices. *Nature.* **497**, 598–602 (2013).

19. L. A. Ponomarenko, R. V. Gorbachev, G. L. Yu, D. C. Elias, R. Jalil, A. A. Patel, A. Mishchenko, A. S. Mayorov, C. R. Woods, J. R. Wallbank, M. Mucha-Kruczynski, B. A. Piot, M. Potemski, I. V. Grigorieva, K. S. Novoselov, F. Guinea, V. I. Fal'ko, A. K. Geim, Cloning of Dirac fermions in graphene superlattices. *Nature.* **497**, 594–597 (2013).

20. R. V. Gorbachev, J. C. W. Song, G. L. Yu, A. V. Kretinin, F. Withers, Y. Cao, A. Mishchenko, I. V Grigorieva, K. S. Novoselov, L. S. Levitov, A. K. Geim, Detecting topological currents in graphene superlattices. *Science.* **346**, 448–451 (2014).

21. Y. Cao, V. Fatemi, S. Fang, K. Watanabe, T. Taniguchi, E. Kaxiras, P. Jarillo-Herrero, Unconventional superconductivity in magic-angle graphene superlattices. *Nature.* **556**, 43–50 (2018).

22. Y. Cao, V. Fatemi, A. Demir, S. Fang, S. L. Tomarken, J. Y. Luo, J. D. Sanchez-Yamagishi, K. Watanabe, T. Taniguchi, E. Kaxiras, R. C. Ashoori, P. Jarillo-Herrero, Correlated insulator behaviour at half-filling in magic-angle graphene superlattices. *Nature*. **556**, 80–84 (2018).

23. A. L. Sharpe, E. J. Fox, A. W. Barnard, J. Finney, K. Watanabe, T. Taniguchi, M. A. Kastner, D. Goldhaber-Gordon, Emergent ferromagnetism near three-quarters filling in twisted bilayer graphene. *Science*. **365**, 605–608 (2019).

24. M. Serlin, C. L. Tschirhart, H. Polshyn, Y. Zhang, J. Zhu, K. Watanabe, T. Taniguchi, L. Balents, A. F. Young, Intrinsic quantized anomalous Hall effect in a moiré heterostructure. *Science*. **367**, 900–903 (2020).

25. L. Li, M. Wu, Binary Compound Bilayer and Multilayer with Vertical Polarizations: Two-Dimensional Ferroelectrics, Multiferroics, and Nanogenerators. *ACS Nano*. **11**, 6382–6388 (2017).

26. N. Marom, J. Bernstein, J. Garel, A. Tkatchenko, E. Joselevich, L. Kronik, O. Hod, Stacking and registry effects in layered materials: The case of hexagonal boron nitride. *Phys. Rev. Lett.* **105**, 046801 (2010).

27. K. Kim, M. Yankowitz, B. Fallahazad, S. Kang, H. C. P. Movva, S. Huang, S. Larentis, C. M. Corbet, T. Taniguchi, K. Watanabe, S. K. Banerjee, B. J. Leroy, E. Tutuc, Van der Waals Heterostructures with High Accuracy Rotational Alignment. *Nano Lett.* **16**, 1989–1995 (2016).

28. Y. Cao, J. Y. Luo, V. Fatemi, S. Fang, J. D. Sanchez-Yamagishi, K. Watanabe, T. Taniguchi, E. Kaxiras, P. Jarillo-Herrero, Superlattice-Induced Insulating States and Valley-Protected Orbits in Twisted Bilayer Graphene. *Phys. Rev. Lett.* **117**, 116804 (2016).

29. R. D. King-Smith, D. Vanderbilt, Theory of polarization of crystalline solids. *Phys. Rev. B*. **47**, 1651–1654 (1993).

30. H. Yoo, R. Engelke, S. Carr, S. Fang, K. Zhang, P. Cazeaux, S. H. Sung, R. Hovden, A. W. Tsen, T. Taniguchi, K. Watanabe, G. C. Yi, M. Kim, M. Luskin, E. B. Tadmor, E. Kaxiras, P. Kim, Atomic and electronic reconstruction at the van der Waals interface in twisted bilayer graphene. *Nat. Mater.* **18**, 448–453 (2019).

31. A. Weston, Y. Zou, V. Enaldiev, A. Summerfield, N. Clark, V. Zólyomi, A. Graham, C. Yelgel, S. Magorrian, M. Zhou, J. Zultak, D. Hopkinson, A. Barinov, T. Bointon, A. Kretinin, N. R. Wilson, P. H. Beton, V. I. Fal'ko, S. J. Haigh, R. Gorbachev, Atomic reconstruction in twisted bilayers of transition metal dichalcogenides, *Nat. Nanotech.* 10.1038/s41565-020-0682-9 (2020).

32. Q. Yang, M. Wu, J. Li, Origin of Two-Dimensional Vertical Ferroelectricity in $WTe_2$ Bilayer and Multilayer. *J. Phys. Chem. Lett.* **9**, 7160–7164 (2018).

33. J. Sung, Y. Zhou, G. Scuri, V. Zólyomi, T. I. Andersen, H. Yoo, D. S. Wild, A. Y. Joe, R. J. Gelly, H. Heo, D. Bérubé, A. M. M. Valdivia, T. Taniguchi, K. Watanabe, M. D. Lukin, P. Kim, V. I. Fal'ko, H. Park, Broken mirror symmetry in excitonic response of reconstructed domains in twisted $MoSe_2$/$MoSe_2$ bilayers. arXiv:2001.01157 [cond-mat.mes-hall] (5 Jan 2020).

**Acknowledgments:** We thank Sergio de la Barrera, Denis Bandurin, Zhiren Zheng, Qiong Ma, Yang Zhang and Liang Fu for fruitful discussions, and Jeong Min Park, and Eric Soriano for experimental support. **Funding:** This research was partially supported by DOE BES grant DE-SC0018935 (early characterization measurements and device nanofabrication), by the Center for the Advancement of Topological Semimetals, an Energy Frontier Research Center funded by the U.S. Department of Energy Office of Science, through the Ames Laboratory under contract DE-AC02-07CH11358 (performance measurements and data analysis), the Army Research Office (early effort towards device nanofabrication) through grant #W911NF1810316, and the Gordon and Betty Moore Foundations EPiQS Initiative through grant GBMF9643 to P.J-H.. This work made use of the Materials Research Science and Engineering Center Shared Experimental Facilities supported by the National Science Foundation (NSF) (Grant No. DMR-0819762). This work was performed in part at the Harvard University Center for Nanoscale Systems (CNS), a member of the National Nanotechnology Coordinated Infrastructure Network (NNCI), which is supported by the National Science Foundation under NSF ECCS award no. 1541959. K.W. and T.T. acknowledge support from the Elemental Strategy Initiative conducted by the MEXT, Japan, Grant Number JPMXP0112101001, JSPS KAKENHI Grant Numbers JP20H00354 and the CREST(JPMJCR15F3). K. Y. acknowledges partial support by JSPS Overseas Research Fellowships. **Author contributions:** K.Y. and P.J-H. conceived the project. K.Y. and X.W. fabricated the devices and performed the measurements. K.W. and T.T. grew the hexagonal boron nitride bulk crystals. K.Y., X.W., and P.J-H. analyzed, interpreted the data, and wrote the manuscript with contributions from all authors. **Competing interests:** The authors declare no competing interests.

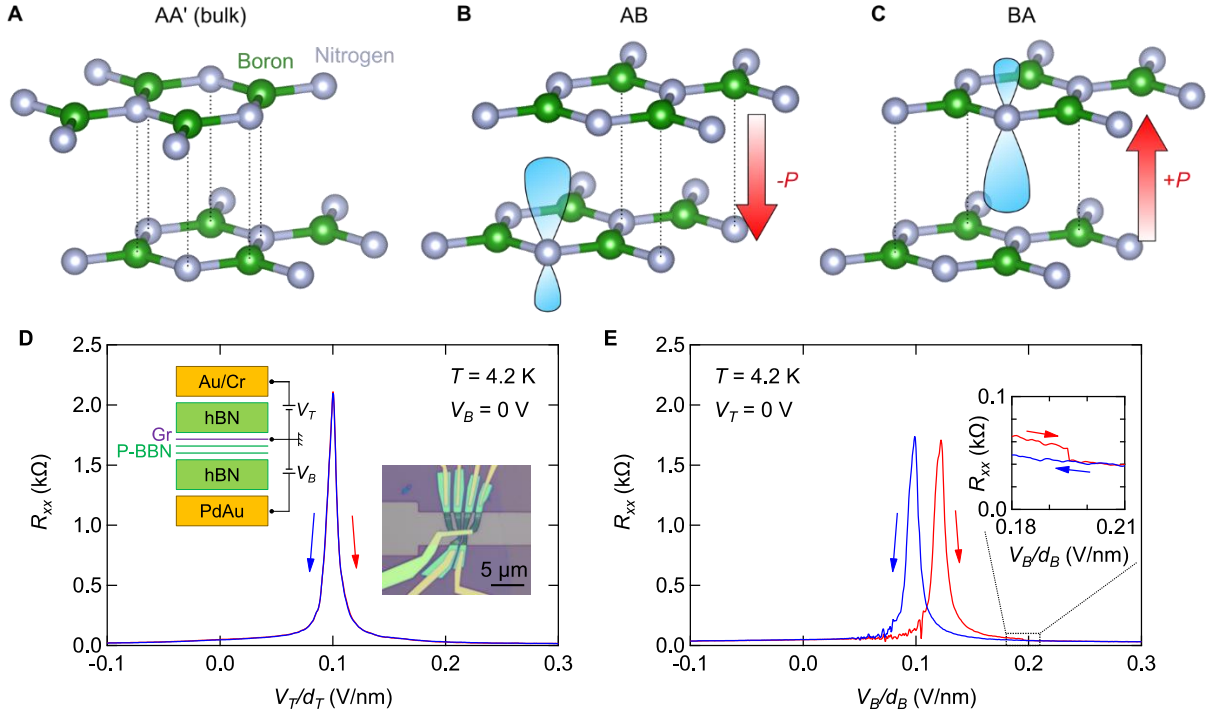

**Fig. 1. Ferroelectricity of AB-stacked bilayer boron nitride.** (**A**) Schematic illustration of the atomic arrangement for AA' stacking, the bulk form of hexagonal boron nitride. Nitrogen and boron atoms are shown in silver and green, respectively. (**B** and **C**) Schematic illustration of the atomic arrangement for AB and BA stacking. The vertical alignment of nitrogen and boron atoms distorts the $2p_z$ orbital of nitrogen (light blue), creating an out-of-plane electric dipole. (**D**) Resistance $R_{xx}$ of graphene for device P1 as a function of $V_T/d_T$, the top gate voltage $V_T$ divided by the thickness of top hBN $d_T$. $V_T/d_T$ is scanned in the backward (forward) direction starting from +0.36 V/nm (−0.36 V/nm) in the blue (red) curve. Note that we only show the relevant scan range around the resistance peak in the figure. The inset on the left shows the schematic device structure. The inset on the right shows an optical micrograph of the device. (**E**) Resistance $R_{xx}$ as a function of $V_B/d_B$, the bottom gate voltage $V_B$ divided by the distance between graphene and bottom gate electrode $d_B$. $V_B/d_B$ is scanned in the backward (forward) direction starting from +0.42 V/nm (−0.42 V/nm) in the blue (red) curve. The inset shows the enlarged plot around 0.20 V/nm.

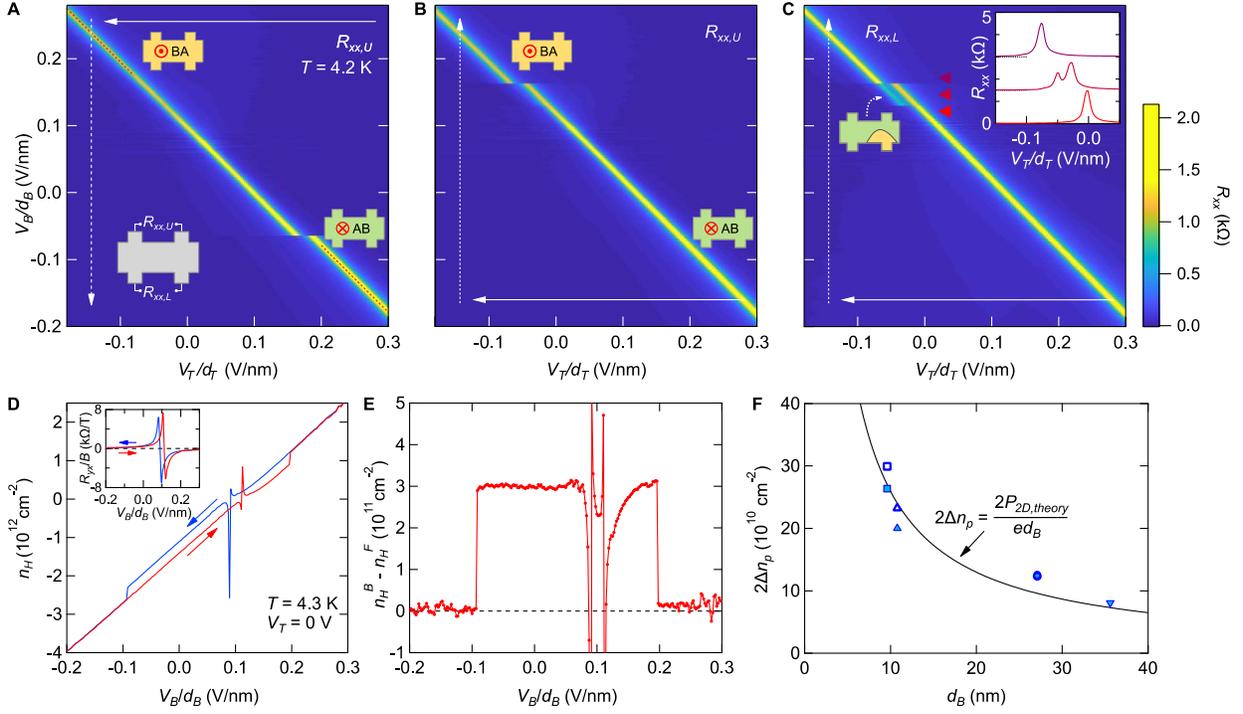

**Fig. 2. Ferroelectricity and polarization in a dual-gated device.** (**A**) Resistance $R_{xx,U}$ measured with the upper voltage contacts of device P1 (as displayed in the inset on the lower left) as a function of $V_B/d_B$ and $V_T/d_T$. We repeatedly scanned $V_T/d_T$ (fast-scan, solid arrow) in the backward direction while gradually changing $V_B/d_B$ (slow-scan, dotted arrow). $V_B/d_B$ is changed in the backward direction starting from +0.42 V/nm. Note that we only show the relevant scan range in the figure. The insets on the upper left and the lower right show the schematic domain configuration and the polarization direction (red). (**B**) The same as **A** with $V_B/d_B$ changed in the forward direction starting from −0.42 V/nm. (**C**) The same as **B** for the resistance $R_{xx,L}$ measured with the lower voltage contacts. The inset on the right shows the line cuts at the fixed $V_B/d_B$ locations indicated by the red triangles. Each curve is offset by 1.5 kΩ for clarity. The inset on the left shows the schematic domain configuration during the ferroelectric switching (fig. S9). (**D**) Hall carrier density $n_H$ measured as a function of $V_B/d_B$. $V_B/d_B$ is scanned in the backward (forward) direction starting from +0.42 V/nm (−0.42 V/nm) in the blue (red) curve. The inset shows the Hall resistance $R_{yx}$ as a function of $V_B/d_B$ under $B = 0.5$ T. (**E**) The difference of Hall carrier density in the backward and forward scan, $n_H^B - n_H^F$. (**F**) Twice the induced carrier density by the polarization of P-BBN, $2\Delta n_P$, plotted against $d_B$ for four devices studied in this work. $2\Delta n_P$ of each device is shown with a different shape; square (P1), triangle (P2), inverse-triangle (T1), and circle (T2). The filled symbols and hollow symbols represent $2\Delta n_P$ estimated from the horizontal shift of the resistance peak and the Hall resistance, respectively. Note that two markers of device T2 almost overlap with each other. The black curve is the theoretical curve calculated from the polarization obtained from Berry phase calculation, $P_{2D,theory}$.

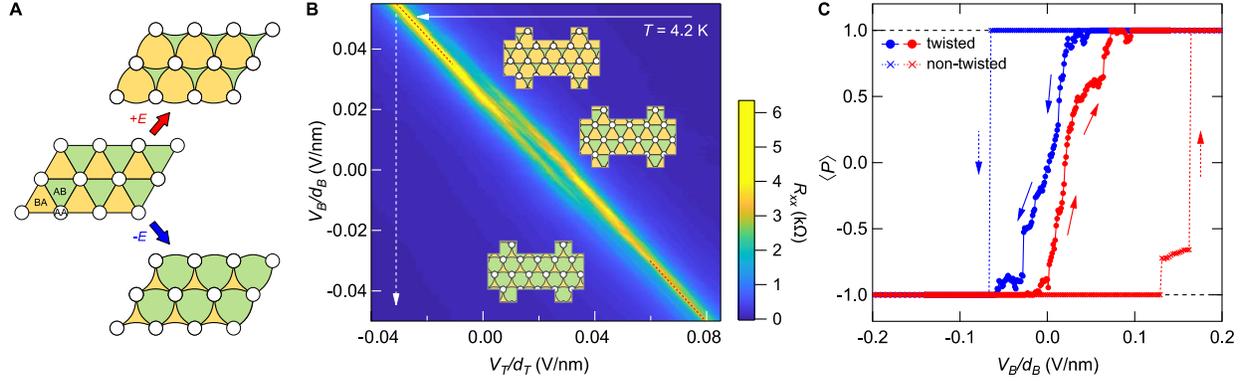

**Fig. 3. Ferroelectric switching in twisted bilayer boron nitride.** (**A**) Schematic illustration of a small-angle twisted bilayer BN after the atomic reconstruction. The reconstruction creates relatively large AB (green) and BA (yellow) domains, with small AA regions (white) and domain walls in between (black). The vertical electric field $\pm E$ is expected to expand or shrink the AB and BA domains. (**B**) Resistance $R_{xx}$ of device T1 as a function of $V_B/d_B$ and $V_T/d_T$. The insets show the schematic illustration of the domain configurations. We repeatedly scanned $V_T/d_T$ (fast-scan, solid arrow) while gradually changing $V_B/d_B$ (slow-scan, dotted arrow). $V_B/d_B$ is changed in the backward direction starting from +0.42 V/nm. (**C**) Spatial average of polarization of bilayer BN, $\langle P \rangle$, estimated from the two-peak fitting as a function of the applied electric field $V_B/d_B$ for a twisted device, T1 (solid lines) and a non-twisted device, P1 (dotted lines). The blue and red curves are backward and forward scans, respectively. $\langle P \rangle$ of device P1 is estimated by taking the average of the polarization measured with the upper voltage contacts and the lower voltage contacts. We expect that small but finite AB (BA) region remains even at $\langle P \rangle = 1\,(-1)$ in the twisted device as depicted in the insets of **B**, although it is too small to be detected in resistance.

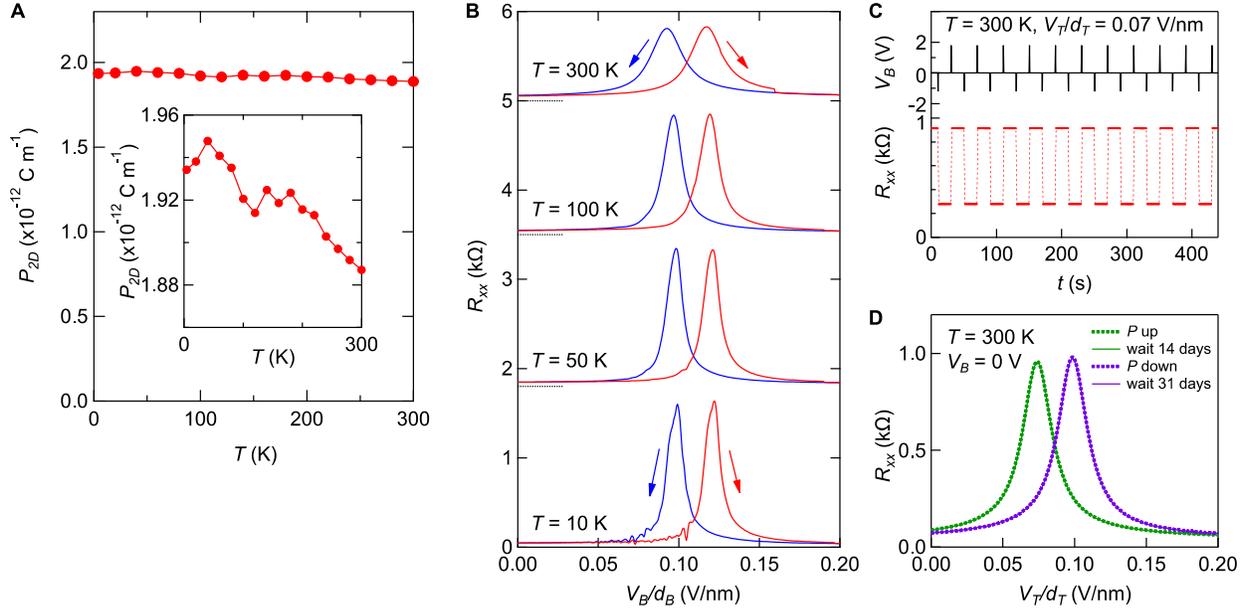

**Fig. 4. Temperature dependence and room-temperature operation.** (**A**) Temperature dependence of the magnitude of the polarization $P_{2D}$ for device P1. The inset shows a zoom-in of the vertical axis. (**B**) Hysteresis of resistance at various temperatures. $V_B/d_B$ is scanned in the backward (forward) direction starting from +0.42 V/nm (−0.42 V/nm) in the blue (red) curve. Each curve is offset for clarity. The offset values are shown in dotted lines. (**C**) Resistance (red curve) after the repeated application of a voltage pulse of $V_B$ = +1.8 V and $V_B$ = −1.2 V (black curve), which corresponds to $V_B/d_B$ = +0.19 V/nm and $V_B/d_B$ = −0.13 V/nm, respectively. The measurement is performed at $T$ = 300 K and $V_T/d_T$ = 0.07 V/nm. (**D**) Stability of polarization at room temperature. $V_T/d_T$ is scanned in the forward direction. The dotted green (purple) curve is measured at $V_B$ = 0 V right after applying $V_B/d_B$ = +0.31 V/nm (−0.26 V/nm) to induce polarization up (down). The solid green curve is measured after applying $V_B/d_B$ = +0.31 V/nm to induce polarization up and then leaving the device at $V_B$ = 0 V at $T$ = 300 K for 14 days. The solid purple curve is measured after applying $V_B/d_B$ = −0.26 V/nm to induce polarization down and then leaving the device at $V_B$ = 0 V at $T$ = 300 K for 31 days. Each of the two curves almost exactly overlaps, showing the robustness of polarization direction for a long period.